%% file: bam_revised.tex
\documentclass[tikz,fleqn,usenatbib]{mnras}

\usepackage{amsmath}
\usepackage{lastpage}
\usepackage{mathptmx}
\usepackage{txfonts}
\usepackage[T1]{fontenc}
\usepackage{ae,aecompl}
\usepackage{tikz}
\usetikzlibrary{matrix, fit, positioning,arrows.meta,arrows, shapes.geometric,backgrounds}

\tikzstyle{startstop}=[rectangle, rounded corners, minimum width=2.7cm, minimum height= 1cm, text centered, draw=black, fill=red!20, text width = 2cm]
\tikzstyle{startstop_nb}=[rectangle, rounded corners, minimum width=2.5cm, minimum height=0.8cm, text centered, draw=black, fill=gray!20, text width = 2cm]
\tikzstyle{io}=[trapezium, trapezium left angle = 70, trapezium right angle =110,minimum width=2cm, minimum height= 1cm, text centered, draw=black, fill=blue!30]
\tikzstyle{io_short}=[rectangle,rounded corners, minimum width=0.2cm, minimum height= 1cm, text centered, draw=black, fill=blue!5, line width = 0.05cm]
\tikzstyle{process}=[rectangle, rounded corners,minimum width=3cm, minimum height= 0.7cm, text centered, draw=black, fill=orange!30]
\tikzstyle{process_DUMMY}=[rectangle, minimum width=0cm, minimum height= 0cm, draw=white, fill=white!0]
\tikzstyle{process_SDUMMY}=[circle, minimum width=0cm, minimum height= 0cm, draw=white, fill=white!0]

\tikzstyle{process_NO_BOX}=[circle, minimum width=0.5cm, minimum height= 0.5cm, text centered, draw=gray, fill=white!0]
\tikzstyle{process_TITLE}=[rectangle, rounded corners, minimum width=1cm, minimum height= 0.4cm, text centered, draw=black, fill=gray!30]

\tikzstyle{nomen}=[rectangle, rounded corners, minimum width=4cm, minimum height= 1cm,  draw=black, fill=yellow!10]
\tikzstyle{process_short}=[rectangle, rounded corners,minimum width=1cm, minimum height=0.5cm, text centered, draw=black, fill=orange!20]
\tikzstyle{decision}=[diamond, rounded corners,minimum width=2cm, minimum height= 0.7cm, text centered, draw=black, fill=green!20]
\tikzstyle{question}=[diamond,rounded corners, minimum width=2.5cm, minimum height= 0.7cm, text centered, draw=black, fill=blue!20]

\tikzstyle{decisionfwd}=[diamond, rounded corners,minimum width=2cm, minimum height= 0.7cm, text centered, draw=black, fill=green!20]
\tikzstyle{decision_short}=[diamond, rounded corners,minimum width=1cm, minimum height= 1cm, text centered, draw=black, fill=green!20]
\tikzstyle{decision_short_input}=[diamond, minimum width=0.8cm, minimum height=0.05cm, text centered, draw=black, fill=gray!20]
\tikzstyle{arrow}=[thick, ->, >= stealth]
\tikzstyle{arrow_d}=[dotted, ->, >= stealth]
\tikzstyle{arrow_new}=[dotted, ->, >= stealth]
\tikzstyle{arrow_pi}=[dashed, ->, >= stealth, line width=0.03cm]
\tikzstyle{arrow_nl}=[dashed, -, >= stealth, line width=0.03cm]
\tikzstyle{arrow_nls}=[thick, -, >= stealth, line width=0.03cm]
\tikzstyle{arrow_new}=[dotted, ->,>= stealth, color=red, line width = 0.02cm]

\usepackage{graphicx}
\usepackage[pdf]{pstricks}
\usepackage{epstopdf}
\usepackage{amssymb}
\usepackage{latexsym}
\usepackage{tikz-cd}
\usepackage{bm}
\newcommand{\aba}{\textcolor{red}}
\usepackage{natbib,twoopt}
\input{newcomands.tex}

\title[Performance of the Bias Assignment Method]{One simulation to have them all: performance of the Bias Assignment Method against $N$-body simulations}
\author[Balaguera-Antol\'{\i}nez et al.]{\parbox{\textwidth}
  {A. Balaguera-Antol\'{\i}nez\thanks{balaguera@iac.es}$^{1,2}$,Francisco-Shu Kitaura\thanks{fkitaura@iac.es}$^{1,2}$, M. Pellejero-Ib{\'a}{\~n}ez$^{3}$,  Martha Lippich$^{4}$, Cheng Zhao$^{5}$,  Ariel G. S\'anchez$^{4}$, Claudio Dalla Vecchia$^{1,2}$, Ra\'ul E. Angulo$^{3,6}$ and Mart\'{\i}n Crocce$^{7}$\\\\}
  \\
  $^{1}$Instituto de Astrof\'{\i}sica de Canarias, s/n, E-38205, La Laguna, Tenerife, Spain\\
  $^{2}$Departamento de Astrof\'{\i}sica, Universidad de La Laguna, E-38206, La Laguna, Tenerife, Spain\\
  $^{3}$Donostia International Physics Centre   (DIPC), Paseo Manuel de Lardizabal 4, 20018 Donostia-San Sebastian, Spain\\
  $^{4}$Max-Planck-Institut f\"ur extraterrestrische Physik, Postfach 1312, Giessenbachstr., 85741 Garching, Germany\\
  $^{5}$Laboratory of Astrophysics, Ecole Polytechnique F\'ed\'erale de Lausanne (EPFL), Observatoire de Sauverny, 1290 Versoix, Switzerland\\
  $^{6}$IKERBASQUE, Basque Foundation for Science, 48013, Bilbao, Spain.\\ 
  $^{7}$Institut de Ci\'encies de l'Espai, IEEC-CSIC, Campus UAB, Carrer de Can Magrans, s/n, 08193 Bellaterra, Barcelona, Spain\\
}

\begin{document}
\label{firstpage}
\pagerange{\pageref{firstpage}--\pageref{lastpage}} 
\maketitle
\begin{abstract}
In this paper we demonstrate that the information encoded in \emph{one} single (sufficiently large) $N$-body simulation can be used to reproduce arbitrary numbers of halo catalogues, using approximated realisations of dark matter density fields with different initial conditions. To this end we use as a reference one realisation (from an ensemble of $300$) of the Minerva $N$-body simulations and the recently published Bias Assignment Method to extract the local and non-local bias linking the halo to the dark matter distribution. We use an approximate (and fast) gravity solver to generate $300$ dark matter density fields from the down-sampled initial conditions of the reference simulation and sample each of these fields using the halo-bias and a kernel, both calibrated from the arbitrarily chosen realisation of the reference simulation. We show that the power spectrum, its variance and the three-point statistics are reproduced within $\sim 2\%$ (up to $k\sim1.0\,h\,{\rm Mpc}^{-1}$),  $\sim 5-10\%$ and $\sim 10\%$, respectively. Using a model for the real space power spectrum (with three free bias parameters), we show that the covariance matrices obtained from our procedure lead to parameter uncertainties that are compatible within $\sim 10\%$ with respect to those derived from the reference covariance matrix, and motivate approaches that can help to reduce these differences to $\sim 1\%$. Our method has the potential to learn from one simulation with moderate volumes and high-mass resolution and extrapolate the information of the bias and the kernel to larger volumes, making it ideal for the construction of mock catalogues for present and forthcoming observational campaigns such as Euclid or DESI.
\end{abstract}
\begin{keywords}
cosmology: -- theory - large-scale structure of Universe 
\end{keywords}
\section{Introduction}
The cosmological volumes explored by current galaxy surveys such as eBOSS \cite[][]{2016AJ....151...44D}, DES \cite[][]{DES}, and the unprecedented volumes expected to be probed by forthcoming experiments such as Euclid \cite[][]{Euclid} and DESI \cite[][]{DESI}, demand precise and accurate covariance matrices for galaxy clustering analysis \cite[e.g.][]{2013PhRvD..88f3537D,2013MNRAS.432.1928T,2014MNRAS.439.2531P,2015MNRAS.454.4326P,2016MNRAS.457..993P,2017MNRAS.472.4935H, 2019MNRAS.487.2701O}. The complex evolution underlying a galaxy distribution and galaxy physical processes (e.g. dark matter non-linear evolution, galaxy bias, baryon effects, redshift space distortions) implies the necessity of detailed $N$-body simulations (in general, cosmological hydro-simulations) to
construct mock catalogues and generate precise covariance matrices, particularly on small scales. Although this represents an accurate approach to asses robust error estimates on cosmological observables, in practice it imposes strong restrictions in terms of computation times and memory requirements. Recently, \citet[][]{2019MNRAS.482.1786L, 2018MNRAS.tmp.2818C} and \citet[][]{2019MNRAS.485.2806B} compared different methods to generate mock catalogues and covariance matrices for different statistical probes, showing the limits and potential of different approaches currently available in the literature. \\ 

The recently published Bias Assignment Method (\texttt{BAM}) by \cite{2019MNRAS.483L..58B} (Paper I hereafter) has been introduced as an alternative path towards the construction of large number of mock galaxy catalogues. The method exploits the idea of mapping the halo distribution onto an \emph{approximated dark matter density field} (\texttt{A-DMDF} hereafter), based on measurements of the halo-bias. Although this approach has been explored by previous works such as \texttt{PThalos} \citep{2002MNRAS.329..629S}, 
\texttt{MoLUSC} \citep{2008ApJ...678..569S}, \texttt{PATCHY} \citep[][]{2014MNRAS.439L..21K}, \texttt{QPM} \citep{2014MNRAS.437.2594W}, \texttt{EZMOCKS} \cite[][]{2015MNRAS.446.2621C} and \texttt{HALOGEN} \citep{2015MNRAS.450.1856A}, the key difference is that \texttt{BAM} does not assume any special analytical form for the halo-bias relation. Instead, the method makes use of the statistical interpretation of the halo-bias \citep[e.g.][]{1999ApJ...520...24D} and extracts it in the form of a parameter-free probability distribution (besides binning choices) from the combination of the \texttt{A-DMDF} and a reference catalogue of tracers, here dark matter haloes (\texttt{DMH} hereafter), obtained from \emph{one} high resolution $N$-body simulation \cite[see e.g.][for a similar approach to create weak lensing maps]{2017ApJ...839...25S}. This allows us to include almost all possible non-linearities and stochastic features present in the halo-bias \citep[see e.g.][]{2015MNRAS.450.1486A,2019arXiv191012452N}. Another key aspect of \texttt{BAM} is that the \texttt{A-DMDF}s are obtained by evolving the initial conditions (\texttt{IC}s hereafter) of the reference $N$-body simulation using approximated gravity solvers such as \texttt{FastPM} \citep{2016MNRAS.463.2273F} or the Augmented Lagrangian  perturbation theory \citep[\texttt{ALPT,}][]{doi:10.1093/mnrasl/slt101},  thus providing a good template for the large-scale distribution of the dark matter density field (\texttt{DMDF} hereafter)
while accounting for the bias introduced by the approximate representation of the \texttt{DMDF} on small scales. This is achieved by means of a multi-dimensional bias relation beyond the univariate PDF, as suggested by \cite{doi:10.1093/mnras/254.2.315} (see also \cite{2013JCAP...11..048L}).

\begin{figure*}
\begin{tikzpicture}[font=\ttfamily\small,node distance=1.45cm]
\hspace{0cm}
\node(start1)[startstop]{IC: INITIAL CONDITIONS \\ $\delta^{i\,\rm ref}_{WN}(\vq)|^{HR}_{N_{\rm cells}}$, \\ $\{\Omega_{\rm m}, \Omega_{\rm b}, \sigma_{8}, n_{s},\cdots\}$ \\ Volume, $M_{\rm dm}$, \\ $P_{\rm lin}(|\vk|)$};

\node(DS1)[decision_short, right of = start1, xshift=0.9cm]{HR$\to$LR};

\draw[arrow](start1)--(DS1);

\node(REFLR)[process_short, right of =DS1, xshift = 0.35cm]{$\delta^{i\,\rm ref}_{\rm WN}(\vq_{i})|^{LR}_{\tilde{N}_{\rm cells}}$};

\draw[arrow](DS1)--(REFLR);

\node(ELMAP)[decision, right of = REFLR, xshift=0.9cm, text width=0.7cm, align=center]{LEMap\\\scriptsize$\vq \to \vr$};

\node(PROD)[process_TITLE, above of= ELMAP, yshift=0.4cm, xshift=-0.5cm]{\large{\texttt{BAM} Calibration}};

\draw[arrow](REFLR)--(ELMAP);

\node(DELTAR)[process_short, right of = ELMAP, xshift = .6cm]{$\delta^{i}_{\rm dm}(\vr_{i})$};

\draw[arrow](ELMAP)--(DELTAR);

\node(APK)[decisionfwd, right of = DELTAR, xshift=0.75cm]{$\mathcal{K}$Conv};

\draw[arrow_pi](DELTAR)--(APK);

\node(RHOC)[process_short, below of = APK, yshift=-0.08cm, xshift=-0.03cm]{{$\tilde{\delta}^{i}_{\alpha\,\rm dm}(\vr_{i})=\delta^{i}_{\alpha\,\rm dm}\otimes \mathcal{K}^{\alpha-1}_{i}$}};

\node(CW)[decision_short, left of = RHOC, xshift=-1.35cm, text width = 0.7cm]{Non-local\\ class.};

\node(RHOCL)[process_short, left of = CW, xshift=-0.65cm]{$\tilde{\Theta}^{i}_{\alpha\,\rm dm}(\vr_{i})$};

\draw[arrow_pi](RHOC)--(CW);

\draw[arrow_pi](APK)--(RHOC);

\node(start2)[decision_short_input, below of = start1, yshift=-0.6cm]{$N$-BS};

\draw[arrow](start1)--(start2);

\node(start3a)[startstop_nb, below of = start2, yshift=0.2cm]{DM Catalogue};

\draw[arrow](start2)--(start3a);

\node(start3b)[decision_short_input, below of = start3a, yshift=0.3cm]{HF};

\draw[arrow](start3a)--(start3b);

\node(start3)[startstop_nb, below of = start3b, yshift=0.3cm]{Halo Cat. $\{ \vr^{i\,\rm ref}_{\rm h} \}$ };

\draw[arrow](start3b)--(start3);

\node(MAS)[decision_short, right of = start3a, xshift=0.45cm]{MAS};

\draw[arrow](start3)-|(MAS);

\node(NH)[process_short, right of = MAS, xshift=0.3cm]{$\{N_{\rm h}^{i\,\rm ref}(\vr )\}|^{LR}_{N_{\rm cells}}$};

\draw[arrow](MAS)--(NH);

\node(BCALC)[decision, right of = NH, xshift =0.7cm]{BiasCalc};

\draw[arrow](NH)--(BCALC);

\draw[arrow_pi](CW)--(RHOCL);

\draw[arrow_pi](RHOCL)--(BCALC);

\node(BIAS)[process_short, right of = BCALC, xshift=0.95cm]{$\mathcal{B}(N_{\rm h}^{i\,\rm ref}|\Theta^{i}_{\alpha \rm dm})$};

\draw[arrow_pi](BCALC)--(BIAS);

\node(SAMPLE)[decision,right of = BIAS, xshift=1.1cm, text width=1.2cm]{ 
Halo \\ \hspace{-0.1cm}Sampling $i\curvearrowleft i(\rm ref)$};

\draw[arrow_pi](RHOC)--(SAMPLE);

\draw[arrow_pi](BIAS)--(SAMPLE);

\node(NM)[process_short, below of = BIAS, yshift=-0.2cm, xshift=0cm]{$N^{i}_{\alpha \rm h}(\tilde{\vr})$}; 

\draw[arrow_pi](SAMPLE)|-(NM);

\node(PKM)[decision_short, below of = BCALC,yshift=-0.2cm,xshift=-1cm] {PSCalc};

\draw[arrow_pi](NM)--(PKM);

\node(PKM2)[process_short, below of = PKM, xshift=1cm]{$P_{\alpha\, \rm h}^{i}(|\vk|)$}; 

\node(PKR2)[process_short, below of = PKM,xshift=-1cm]{$P_{\rm h}^{i\,\rm ref}(|\vk|)$}; 

\draw[arrow](PKM)--(PKR2);

\draw[arrow](NH)|-(PKM);

\draw[arrow_pi](PKM)--(PKM2);

\node(RATIO)[decision, below of = PKR2, xshift=1.cm, align=center]{$\mathcal{T}$Calc};

\node(RATIO2)[process_short, below of = RATIO, yshift=-0.3cm, text width=1.cm]{$\mathcal{T}^{\alpha}_{i}(|\vk|)$};

\draw[arrow_pi](RATIO)--(RATIO2);

\node(RESIDUALS)[question, below of = RATIO2, yshift=-0.5cm, text width=1.6cm]{\hspace{-0.1cm}Residuals\\ $<1\%$?};

\draw[arrow_pi](RATIO2)--(RESIDUALS);

\node(KERNELC2)[decision, left of = RATIO, xshift=4.0cm,  align=center]{$\mathcal{K}_{\alpha}$Update};

\node(ANSN)[process_NO_BOX, right of = RATIO2, xshift=-0.1cm, yshift=0cm]{No};
\draw[arrow_nl](RESIDUALS)--(ANSN);

\draw[arrow_pi](ANSN)--(KERNELC2);


\node(ANS)[process_NO_BOX, right of = RESIDUALS, xshift=1.2cm, yshift=-0.cm]{Yes};

\draw[arrow_nls](RESIDUALS)--(ANS);

\node(YES)[startstop, right of = RESIDUALS, xshift=4.5cm, yshift=0cm]{
\texttt{BAM} Output: \\
Bias:\\
$\mathcal{B}_{\alpha}(N_{\rm h}^{i\,\rm ref}| \Theta^{i}_{\rm dm})$\\
Kernel:\\ $\mathcal{K}^{\alpha}_{i}(|\vk|)$
};

\draw[arrow](ANS)--(YES);


\draw[arrow](PKR2)--(RATIO);
\draw[arrow_pi](PKM2)--(RATIO);



\node(KK)[process_short, right of = KERNELC2, xshift=0.7cm]{$\mathcal{K}^{\alpha}_{i}(|\vk|)$}; 
\draw[arrow_pi](KERNELC2)--(KK);

\node(ANSIT)[process_NO_BOX, right of = KK, xshift=0.4cm, yshift=-0.cm]{\small{$\alpha\to \alpha+1$}};
\draw[arrow_pi](KK)--(ANSIT);

\node(DUMMYK)[process_DUMMY, right of = ANSIT, xshift=-0.49cm]{}; 

\draw[arrow_nl](ANSIT)--(DUMMYK);

\draw[arrow_pi](DUMMYK)|-(APK);

\node(DUMMY)[process_DUMMY, above of = DELTAR, yshift=-2cm]{};

\node(LRJ)[startstop, right of = DUMMY, xshift=4.2cm, yshift=0.6cm, text width= 2.5cm ]{
Input from \texttt{BAM}: \\
Bias $\mathcal{B}(N_{\rm h}^{i\,\rm ref}| \Theta^{i}_{\rm dm})$\\
Kernel $\mathcal{K}_{i}$\\
IC :\\
$\delta^{j}_{WN}(\vq)|^{LR}_{N_{\rm cells}}$ \\ $j=1,\cdots, N_{\rm samples}$
};
\node(PROD)[process_TITLE, above of= LRJ, yshift=0.4cm,xshift=0.05cm]{\large Mock Production};

\node(ELMAP2)[decision, below  of = LRJ, yshift=-0.8cm, text width=0.7cm, align=center]{LEMap$\,\,$  \\\scriptsize$\vq \to \vr$};

\draw[arrow_new](LRJ)--(ELMAP2);

\node(NEWDM)[process_short, below of = ELMAP2, yshift=0.cm]{$\delta^{j}_{\rm dm}(\vr)$}; 
\draw[arrow_new](ELMAP2)--(NEWDM);

\node(APK2)[decisionfwd, below of  = NEWDM, yshift=0.1cm]{$\mathcal{K}$Conv};

\draw[arrow_new](NEWDM)--(APK2);

\node(RHOC2)[process_short, below of = APK2, yshift=0.1cm]{{$\tilde{\delta}^{j(i)}_{\rm dm}(\vr)=\delta^{j}_{\rm dm}\otimes \mathcal{K}_{i}$}};
\draw[arrow_new](APK2)--(RHOC2);


\node(CW2)[decision_short, below of = RHOC2, yshift=-0.2cm,text width = 0.7cm]{Non-local\\ class.};

\draw[arrow_new](RHOC2)--(CW2);

\node(RHOCLP)[process_short, below of = CW2, yshift=-0.15cm]{$\Theta^{j(i)}_{\rm dm}(\vr)$};

\draw[arrow_new](CW2)--(RHOCLP);

\node(SAMPLE2)[decision,below of = RHOCLP, xshift=0cm, yshift=-0.4cm, text width=1.2cm]{ 
Halo \\ \hspace{-0.1cm}Sampling $j\curvearrowleft i(\rm ref)$};

\draw[arrow_new](RHOCLP)--(SAMPLE2);

\node(MOCK)[startstop, below of = SAMPLE2, yshift=-0.9cm]{
Ouput: Halo number counts\\
$N^{j(i\,\rm ref)}_{\rm h}(\vr)$ \\ $j=1,\cdots, N_{\rm samples}$
}; 

\draw[arrow_new](SAMPLE2)--(MOCK);

\node(NOMEN)[nomen, below of = start3, xshift=0.1cm, yshift=-3.5cm, text width=3.5cm]
{
\aba{BiasCalc}: Measurement of halo-bias $\mathcal{B}$\\
\aba{HF}: Halo Finder\\
\aba{Halo Sampling $j\curvearrowleft i(\rm ref)$}: \\$N^{j}_{h}\curvearrowleft \mathcal{B}(N^{i\, \rm ref}_{\rm h}| \Theta^{j}_{\rm dm})$

\aba{$\mathcal{K}$Conv}: Convolution with \texttt{BAM} kernel \\
\aba{$\mathcal{K}_{\alpha}$Update}: Update \texttt{BAM} kernel at the current iteration $\alpha$\\
\aba{LEMap}: Lagrangian ($\vq$) to Eulerian ($\vr$) mapping \\
\aba{LR (HR)}: Low (High) Resolution \\
\aba{MAS}: Mass Assignment Scheme\\
\aba{$N$-BS}: $N$-body solver\\
\aba{PSCalc}: Measurement of Power Spectrum $P(|\vk|)$\\
\aba{$\mathcal{T}$Calc}: Calculate transfer function $\mathcal{T}(|\vk|)=P_{\rm h}^{i\, \rm ref}(|\vk|)/P_{\alpha\, \rm h}^{i}(|\vk|)$\\
\aba{$\Theta_{\rm dm}$}: Set of local \& Non-local properties of the DM field 
};


\node(corner)[process_DUMMY, right of = RESIDUALS, xshift=6cm, yshift=-3.5cm]{};
\draw[dotted, thick] (start1.north) + (-2., 1.1) rectangle (corner.east) + (7.0,-1.0);

\node(corner2)[process_DUMMY, right of = MOCK, xshift=0cm, yshift=-1.3cm]{};
\draw[dotted, thick] (LRJ.north) + (-1.5, 1.0) rectangle (corner2.east) + (-0.5,0.3);

\end{tikzpicture}
\vspace{-1.2cm}
\caption{Flow-chart representing the operations performed within \texttt{BAM}. On the left-hand side box (\texttt{\texttt{BAM} Calibration}) we show the iterative process developed in order to calibrate the kernel $\mathcal{K}$ and the halo-bias using \emph{one} realisation of a halo catalogue from the reference simulation, $\{\vr^{i\,\rm ref}_{\rm h}\}$, in combination with a dark matter density field obtained from the approximated gravity solver with the corresponding initial conditions $\delta^{i\,\rm ref}_{\rm WN}(\vq)$. Solid line arrows denote the process involving the reference $N$-body simulation, performed once. Dotted-line arrows denote the iterative process, which finishes once the relative residuals (see text) are below $1\%$. The index $\alpha$ is assigned to quantities being updated within the iterative process, while $i$ identifies a white noise from the reference simulation. For the first iteration the kernel is $\mathcal{K}^{\alpha=0}_{i}(|\vk|)=1$. The right hand side box (\texttt{Mock Production})  depicts the process followed to generate mock halo density fields based on the outputs of \texttt{BAM}.} 
  \label{bam_scheme}
\end{figure*}
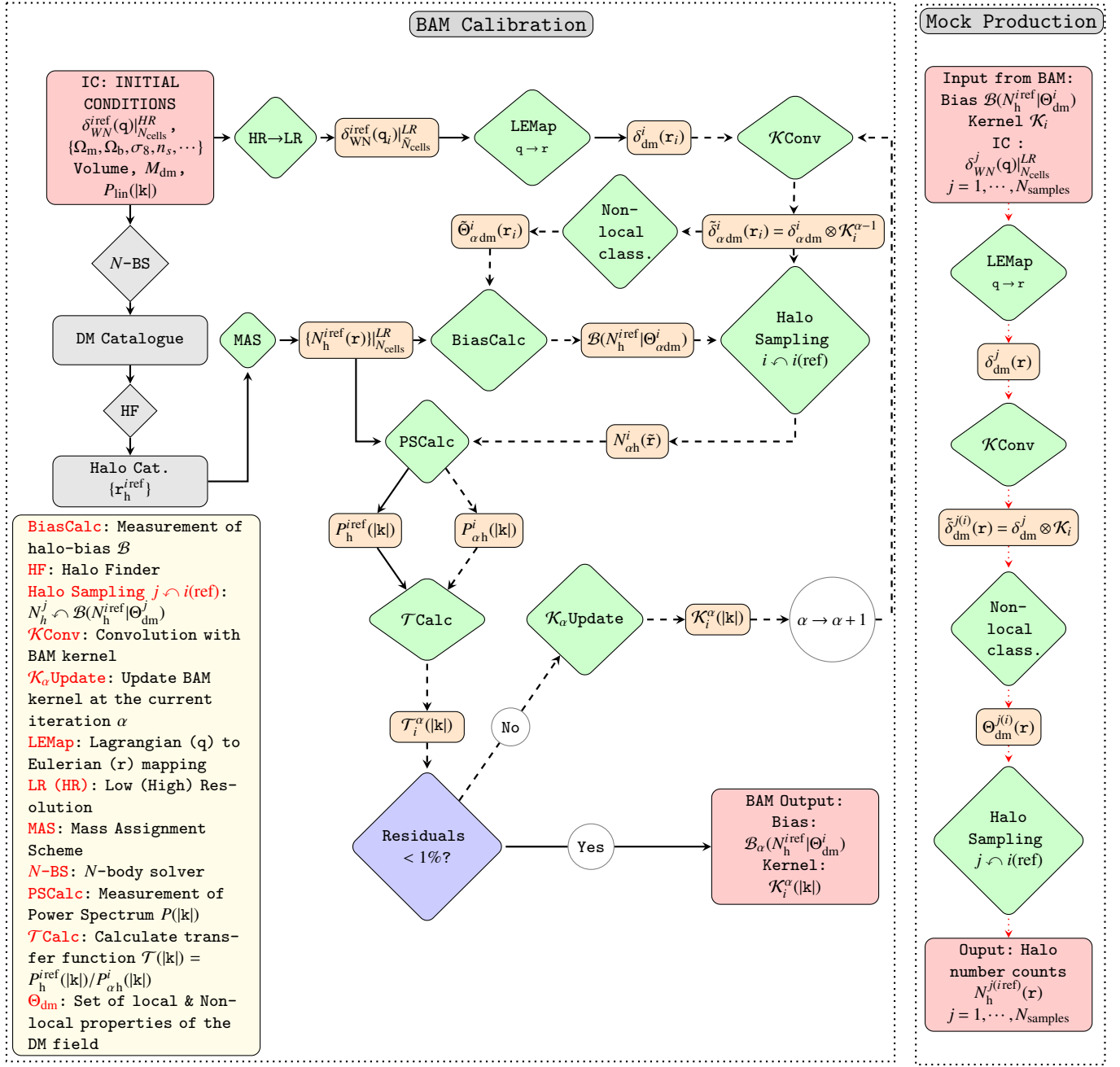

The \texttt{BAM} approach has been envisaged to generate mock catalogues with a power spectrum that replicates, to per-cent precision, that of the high-resolution halo distribution used as a reference. This, together with the description of the method, has been shown and discussed in Paper I. In this article we want to go deeper into the validation of \texttt{BAM}, by assessing its capability to generate independent realisations of mock catalogues based on the calibration of the halo-bias obtained \emph{from only one realisation of the reference simulation.}  In particular, we quantify such capability by comparing relevant summary statistics in Fourier space (such as the two-point statistics by means of the halo power spectrum, and the three-point represented by the halo bispectrum) and by testing the performance of the covariance matrix of the power spectrum from these catalogues against that of an ensemble of reference $N-$body simulations in the framework of a likelihood analysis. We show that our method can generate covariance matrices that yield errors in model parameters, which  are compatible within $\sim 1\%$ with respect to  those derived from a set of reference $N-$body simulations, giving \texttt{BAM} the potential to become one of the standard techniques to generate mock catalogues for large-scale structure analysis.

The outline of this paper is as follows. In \S~\ref{sec:meth} we describe the procedure to generate ensembles of halo density fields with \texttt{BAM}. In \S~\ref{sec:results} we show the comparison of the statistical properties of a set of mocks created with \texttt{BAM} against that from the reference ensemble. In particular, we check the behaviour of the mocks in terms of the two point statistics in Fourier space, as well as at the level of a likelihood analysis based on the power spectrum. We finish with discussion and conclusions.



\section{Methodology}\label{sec:meth}
This section presents a deeper exposition of the procedures followed within \texttt{BAM}, already introduced in Paper I. In order to facilitate the understanding of the procedure, we have depicted the main steps in the flow-chart of Fig.~\ref{bam_scheme}, whose notation we will follow along this paper.

\subsection{Inputs}\label{sec:inputs}
The main inputs of \texttt{BAM} are, on one side, one realisation of the \texttt{IC}s of a high-resolution $N$-body simulation and, on the other side, the halo catalogue corresponding to that particular realisation of the reference simulation. As the reference we have adopted the Minerva simulations \citep{2016MNRAS.457.1577G}, that consists in a set of $N_{\rm sim}=300$ realisations of dark matter and dark matter halo distributions, each embedded in a cubic box of $L_{\rm box}=1.5h^{-1}$ Gpc side. Dark matter haloes are identified with a standard Friends-of-Friends (FoF) algorithm at redshift $z = 1$, and subjected to an unbinding procedure \citep{2001MNRAS.328..726S}, in which particles with positive total energy are removed and haloes artificially
linked by FoF are separated. The minimum halo mass is $\sim 2.7\times 10^{12} h^{-1}\,M_{\odot}$, accounted with a minimum of ten dark matter particles. Although naively one might expect inaccuracies produced by
this low threshold \cite[see. e.g][]{2015MNRAS.453.1513C}, these are largely suppressed since we make
comparisons with the reference using the same initial white noise
field.

\subsection{Calibration}\label{sec:calibration}
The calibration process in \texttt{BAM} aims at delivering two products, namely, a kernel, and a halo-bias, with which a set of halo mock catalogues are to be constructed. The role of the kernel is to correct for any effective large scale contributions which may come from non-local bias dependencies not modeled in \texttt{BAM} \citep[see e.g.][]{2009JCAP...08..020M}, on the one hand, and on the other hand, to correct for any aliasing effects caused by the representation of the \texttt{DMF} and the halo distribution on a mesh with respect to the original halo finding algorithm used to construct the reference catalogue.

As depicted in Fig.~\ref{bam_scheme}, the calibration procedure starts with a down-sampling of the \texttt{IC}s from a high resolution white noise into a low resolution version thereof. For the current set-up we have transformed the \texttt{IC}s from a $1000^{3}$ into a $500^{3}$ resolution. This new white noise is transformed to Fourier space and multiplied by the squared-root of the linear matter power spectrum in spherical shells, $P_{\rm lin}(|\vk|)$, evaluated at the initial redshift of the simulations. The resulting field is evolved up to $z=1$ to generate the \texttt{A-DMDF} $\delta_{\rm dm}^{i}(\vr)$. This is achieved by using \texttt{ALPT} \citep{doi:10.1093/mnrasl/slt101} combined with the phase-space mapping technique \citep[][]{2012MNRAS.427...61A,Hahn:2013aa}, the latter being proven to be a key ingredient to obtain a reasonable three-point statistics of the mock haloes within \texttt{BAM}, at least for the halo-mass resolution of our reference simulation 
\citep[see e.g.][]{2019arXiv191013164P}. The \texttt{A-DMDF} is generated using the cloud-in-cell mass assignment scheme.

The next step consists in the measurement of the halo-bias and the so-called  \texttt{BAM}-kernel $\mathcal{K}$. Starting with the halo-bias, \texttt{BAM} adopts its stochastic interpretation \citep[see e.g.][]{1999ApJ...520...24D,2000ApJ...540...62S} and measures the probability of finding a number of \texttt{DMH}, $N_{\rm h}$, within the volume of a low-resolution cell, conditional to the set of local and non-local properties of the \texttt{A-DMDF} evaluated at that particular cell. We denote the latter set of $\mathcal{N}_{\rm p}$ properties as $\Theta_{\rm dm}$. As exposed in Paper I, \texttt{BAM} not only considers the local dark matter density as the main halo-bias driver \cite[see e.g][for a recent analysis on the main dependencies of halo-bias]{2019MNRAS.482.1900H}, but uses also non-local information of the \texttt{A-DMDF}. Such non-local properties are represented, on one side, by the cosmic-web classification (CWC) based on the analysis of the eigenvalues of the tidal field of the of \texttt{A-DMDF} \citep[see e.g.][]{2007MNRAS.375..489H, 2018MNRAS.476.3631P}. On the other hand, a second non-local property is represented by the mass of collapsing regions, $M_{k}$, identified using a friend-of-friend algorithm joining cells classified as knots \citep[see e.g. ][]{2015MNRAS.451.4266Z}. For this work we implement $\mathcal{N}_{p}=3$ as follows: 
\begin{equation}\label{list}
\Theta_{\rm dm} \equiv \{ \log_{10}(2+\delta_{\rm dm}), {\rm CWC}, M_{K} \},
\end{equation}
where $\delta_{\rm dm}$ denotes the dark-matter overdensity.\footnote{The set-up of \texttt{BAM} for this work implements $200$ bins for the dark matter density field, $200$ bins for the mass of collapsing regions and a cosmic-web classification made by three components, viz, knots, filaments and the set sheets plus voids.} Explicit third order (or higher) non-local dependencies have not been implemented yet in our method. Although the effects of these elements are absorbed by the kernel, their inclusion could improve the the accuracy and precision of the bispectrum, which, given our reference simulation, still displays $5-10\%$ deviations. \footnote{We note that in \citet[][]{2019arXiv191013164P} it is found that choosing a low mass cut of haloes based on high resolution $N$-body simulations the bispectrum is very well fit, as the galaxy distribution yields a better representation of the dark matter field.}

The halo-bias is measured in \texttt{BAM} as a multi-dimensional histogram representing the joint probability distribution of the number of haloes $N_{\rm h}$ in one cell and the aforementioned properties of the underlying \texttt{DMDF}. We then normalize by the distribution of such properties in order to obtain a conditional distribution function. This can be mathematically expressed as
\be\label{eq:bias}
\mathcal{B}(N_{\rm h}|\Theta_{\rm dm})=\frac{\sum_{i=1}^{N_{\rm cells}}\mathbf{1}_{N_{h}}(N_{\rm h}(\vr_{i}))\prod_{\kappa=1}^{\mathcal{N}_{\rm p}}\mathbf{1}_{\gamma_{\kappa}}(\{\Theta_{\rm dm}(\vr_{i})\}_{\kappa})}{\sum_{i=1}^{N_{\rm cells}}\prod_{\kappa=1}^{\mathcal{N}_{\rm p}}\mathbf{1}_{\gamma_{\kappa}}(\{\Theta_{\rm dm}(\vr_{i})\}_{\kappa})},
\ee
where $\gamma_{\ell}\equiv [\{\Theta_{\rm dm}\}_{\ell}-\Delta_{\ell}/2,\{\Theta_{\rm dm}\}_{\ell}+\Delta_{\ell}/2 )$ represents the set of bins (of width $\Delta_{\ell}$) defined for the $\ell$-th property (listed in Eq.~(\ref{list})) of the \texttt{A-DMDF}, with $\mathbf{1}_{A}(x)$ denoting the \emph{indication function}, $\mathbf{1}_{A}(x)=1$ if $x\in A$, $0$ otherwise. Note that the conditional probability distribution $\mathcal{B}$ carries no information on the phases of the density fields. The methodology represented by Eq.~(\ref{eq:bias}) is an approximation to the true underlying halo-bias, and ignores key aspects thereof, such as the effects of the mass assignment and correlation between pairs in different property-bins, among others. The effects of such missing aspects in the measurement of the bias are nevertheless accounted for within the iterative process, as previously pointed out.

\begin{figure}
\hspace{-0.2cm}
\includegraphics[width=9.5cm, height=7.5cm]{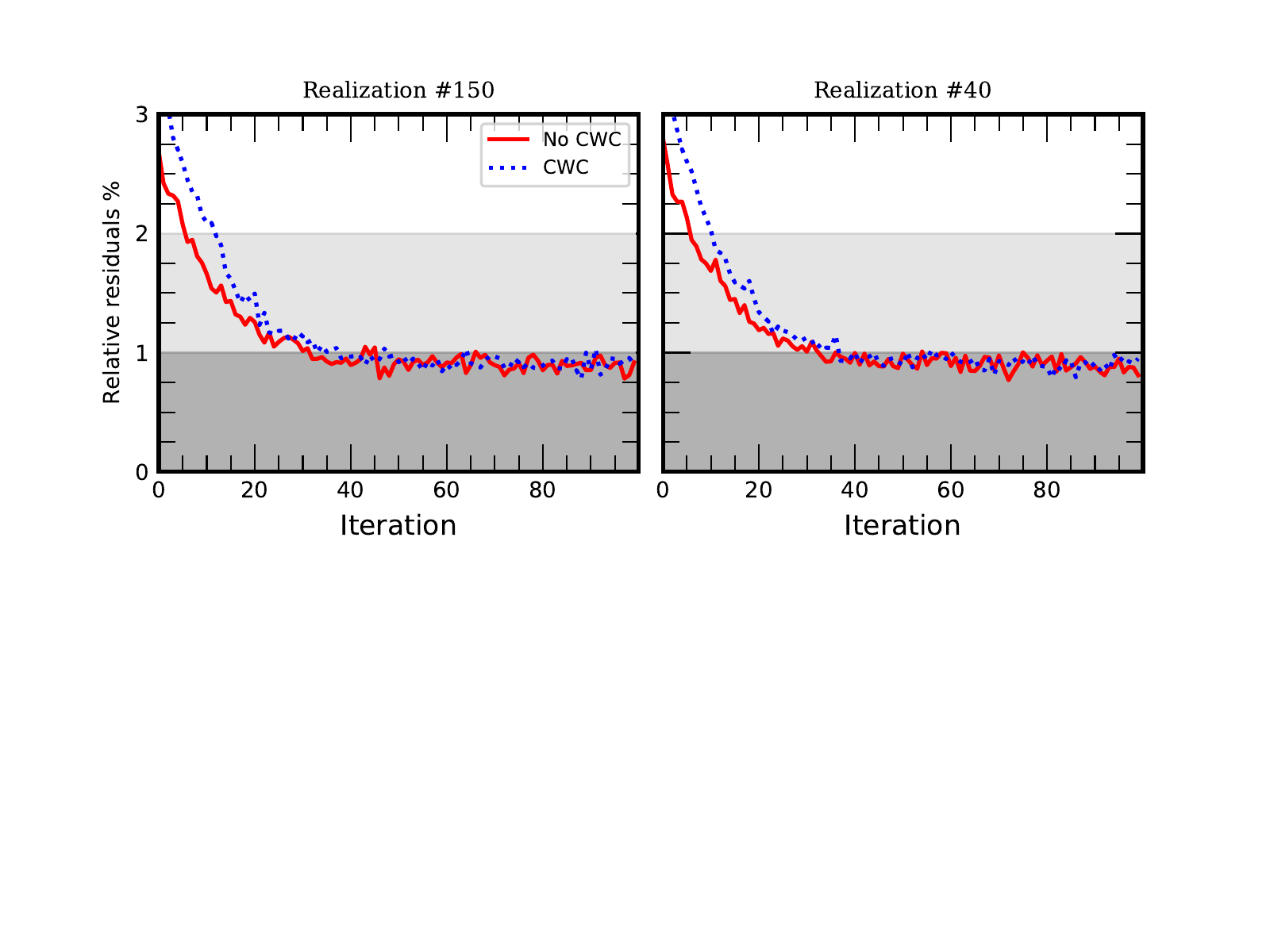}
\vspace{-3.5cm}
\caption{Relative residuals ($\sum_{\kappa}|T_{i}^{\alpha}(k_{\kappa}) -1 |/N_{\rm F}$ see \S~\ref{sec:calibration}) for realisations $\#150$ (on the left) and $\#40$ (on the right) demonstrating the fast convergence of BAM after about 30 iterations. Dashed (solid)  lines represent calculations with (w/o) cosmic-web classification. Light and dark shadowed areas represent the $2\%$ and $1\%$ deviations, respectively. } 
\label{bam_iterations}
\end{figure}

Let us now consider the iterative process in \texttt{BAM}. At the $\alpha$-th iteration, \texttt{BAM} measures the bias $\mathcal{B}(N_{\rm h}^{i\, \rm ref}|\Theta_{\alpha \rm dm}^{i})$ using Eq.~(\ref{eq:bias}), where
$N_{\rm h}^{i\, \rm ref}$ denotes the halo number counts in cells (i.e, using the nearest-grid-point mass assignment scheme) from the high-resolution halo catalogue
\footnote{In this study we have performed the clustering analysis with an NGP mass assignment scheme for both the reference N-body based and the \texttt{BAM} halo catalogs. Performing an analysis with a different mass assignment scheme, such as CIC, could lower the accuracy of the power spectrum to a lower $k$ \citep[see e.g][]{2015MNRAS.450.1836K}. We will investigate this is detail in a forthcoming work.}, and 
\be\label{eq:conv}
\tilde{\delta}_{\alpha \rm dm}^{i}(\vr)\equiv \lp  \mathcal{K}^{\alpha-1}_{i}\otimes \delta_{\rm dm}^{i}\rp(\vr)
\ee
is the convolution of the \texttt{DMDF} with the kernel (in the first iteration, e.g, $\alpha=1$, the \texttt{BAM}-kernel in configuration space is a Dirac's delta function). With the halo-bias at hand, a \texttt{HDF} (strictly speaking, number counts on a grid) $N_{\alpha \rm  h}^{i}(\vr)$ is obtained by statistically sampling the density field $\tilde{\delta}_{\alpha \rm dm}^{i}$ as
\be\label{eq:sam1}
\{N_{\alpha \rm  h}^{i}(\vr)\}  \curvearrowleft \mathcal{B}\lp N_{\rm h}^{i\,\rm ref}    \, \mid \, \Theta_{\rm dm}^{i} =  \Theta_{\alpha \rm dm}^{i}(\vr) \rp,
\ee
after which its power spectrum $P_{i\,\rm h}^{\alpha}(k)$, measured in spherical shells characterised by a wave number $k\equiv |\vk|$,
is used to compute a phase-independent transfer function, \be \label{eq:transfer0}
\mathcal{T}_{i}^{\alpha}(k)\equiv 
\frac{P_{\rm h}^{i\,\rm ref}(k)}{P^{\alpha}_{i\,\rm h}(k)},
\ee
where $P_{\rm h}^{i\,\rm ref}(k)$ denotes the power spectrum of the reference halo catalogue. The sampling procedure of Eq.~(\ref{eq:sam1}) is performed such that the new \texttt{HDF} not only contains the same number of objects as the reference, but also shares its number-count statistics. 

For each spherical shell in Fourier space, \texttt{BAM} implements a Metropolis-Hasting (MH) algorithm to accept ($MH=1$) or reject ($MH=0$) the corresponding value of the transfer function in Eq.~(\ref{eq:transfer0}), using as metric, the quadratic difference of the mock and reference power spectrum in units of the variance thereof, assumed to be a Gaussian \cite[e.g.][]{1994ApJ...426...23F}. \texttt{BAM} defines a set of weights $\omega_{i}^{\alpha}$ according to the outputs of the MH criteria as:
\be \label{eq:transfer}
\omega^{(\alpha)}_{i}(k)\equiv 
\begin{cases}
\mathcal{T}_{i}^{\alpha}(k) & {\text{if $MH=1$}} \\
1 & {\text{if $MH=0$,}}\\
\end{cases}
\ee
and generates an updated version of the \texttt{BAM}-kernel as a product of the weights at the current step with those from the preceding iterations:  
\be \label{eq:kernel}
\mathcal{K}_{i}^{(\alpha)}(k)=\prod_{\ell=1}^{\ell=\alpha}\omega^{(\ell)}_{i}(k)
\ee
(with $\mathcal{K}_{i}^{(\alpha=0)}(k)=1$). This kernel is applied to the \texttt{A-DMDF}, following Eq.~(\ref{eq:conv}) to start again with the procedure, as illustrated in Fig.~\ref{bam_scheme}. 

\begin{figure}
\vspace{-0.4cm}
\hspace{-0.7cm}
\includegraphics[width=17cm]{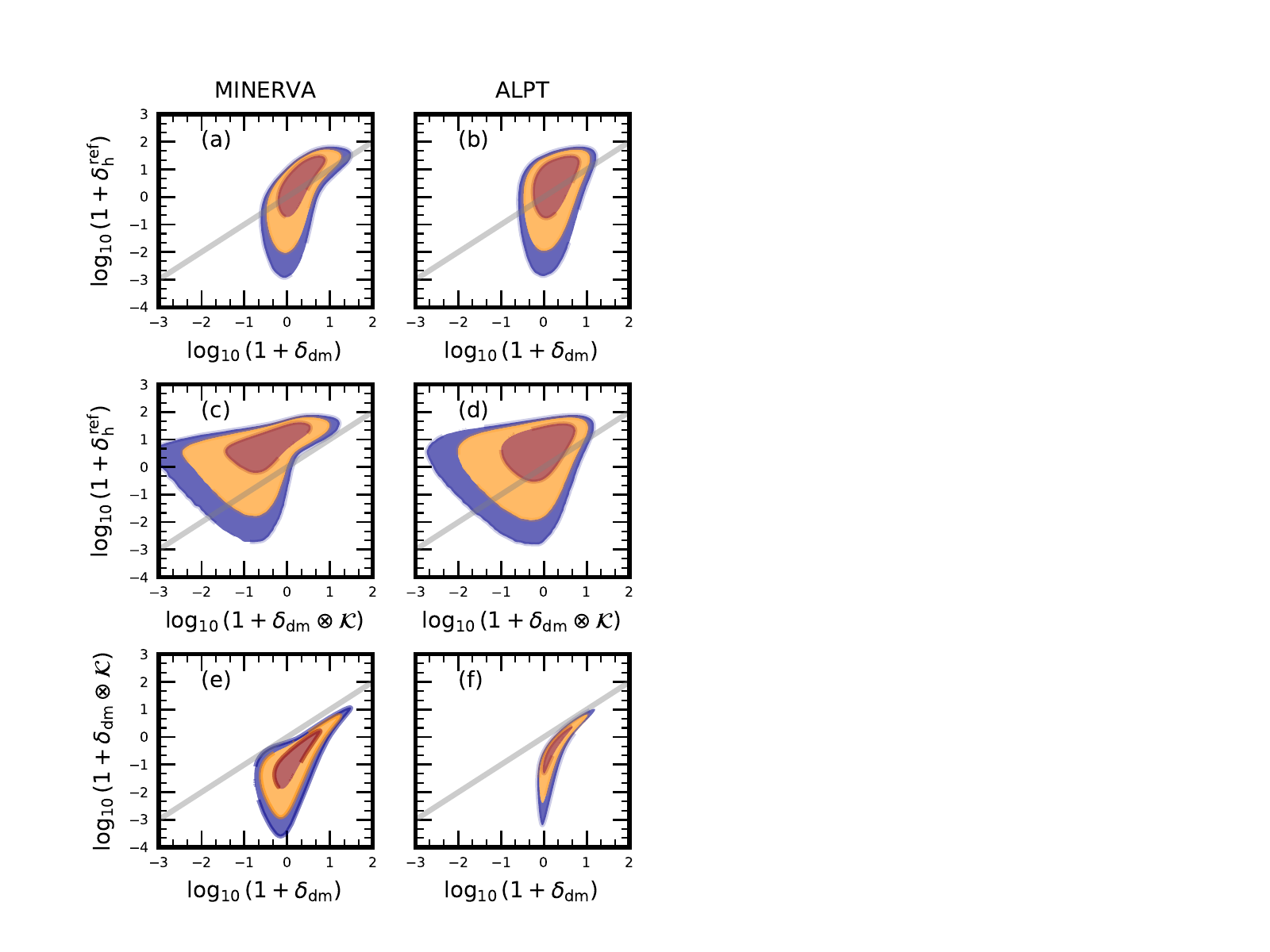}
\vspace{-0.9cm}
\caption{Contours of joint probability distribution in the dark matter - halo plane, from one realisation of the reference simulation, denoting the regions containing the $68\%$ (red), $98\%$ (yellow) and $99\%$ (purple) of the total number of cells. Panel (a) shows the bias measured from a reference halo catalogue using its corresponding dark matter catalogue. Panel (b) shows the bias obtained from the reference halo catalogue with the approximated dark matter field obtained from \texttt{ALPT}. Panel (c) and (d) show the bias from the reference halo catalogue using the approximated dark matter density field convolved with the \texttt{BAM} kernel using the reference \texttt{DMDF} and the \texttt{ALPT} respectively. Panels (e) and (f) show the relation between the approximated density field and its convolution with the \texttt{BAM}  kernel. In all panels the solid line denotes unity bias. Note that the kernels acting in the different columns are different, since these have been calibrated with the different DM fields. For visual purposes, all density fields are obtained using cloud-in-cell interpolation scheme.} 
\label{bam_joints}
\end{figure}

\begin{figure}
\hspace{-0.5cm}
\includegraphics[width=16cm]{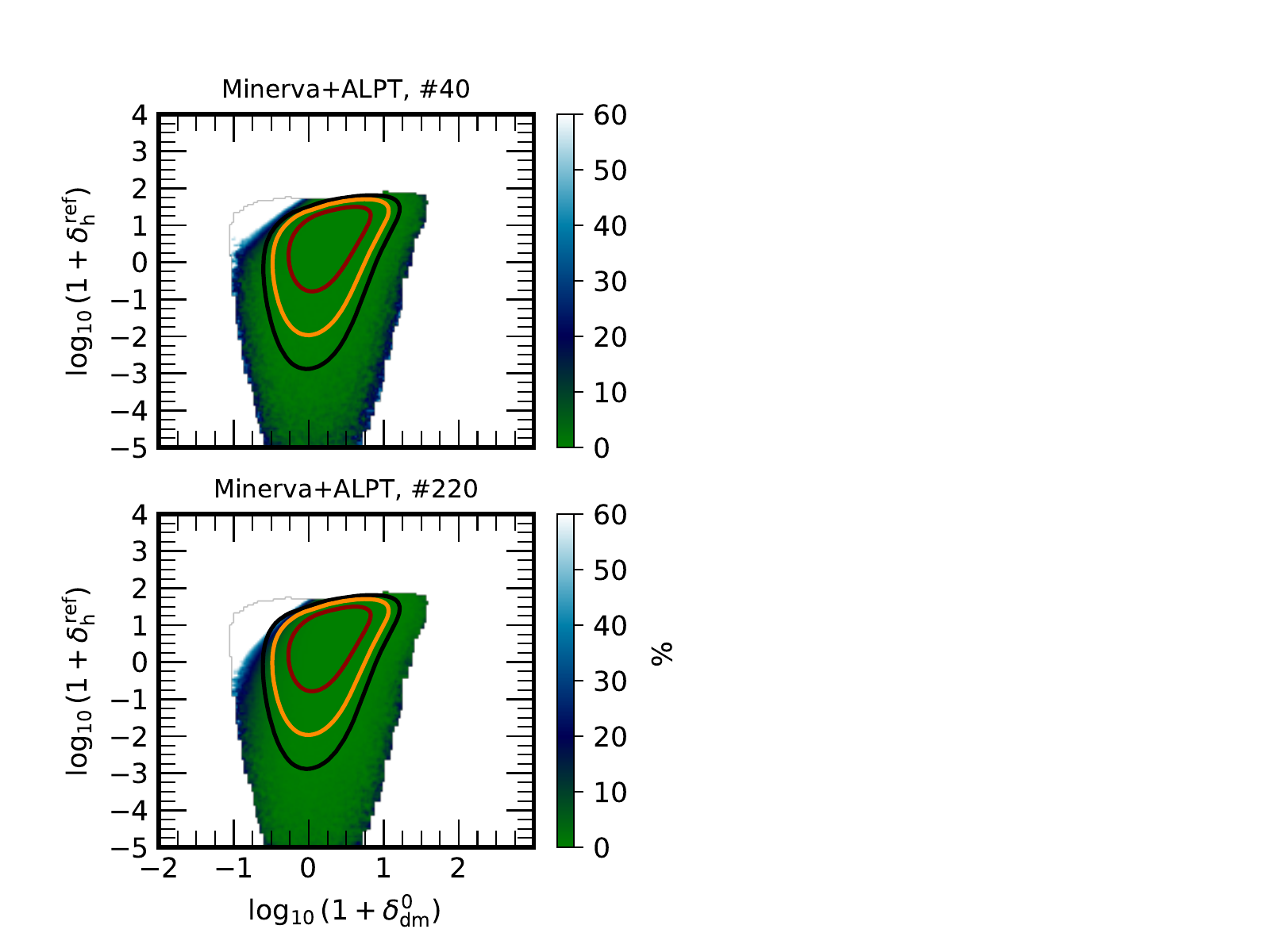}
\vspace{-0.5cm}
\caption{Percent deviation of the joint distribution $\mathcal{B}(N_{\rm h},\delta_{\rm dm})$ measured from two  particular realisations of the ensemble  $300$ Minerva+\texttt{ALPT}, with respect to the mean of the same quantity obtained from that ensemble. The solid lines in all panels denote contours of joint probability distribution containing the $68\%$ (red), $98\%$ (yellow) and $99\%$ (black) of the total number of cells for each individual calibration. Differences with respect to the mean are of the order of $\leq 5\%$ in the regions of the $N_{h}-\delta_{\rm dm}$ plane containing  $\sim 98\%$ of the total number of cells. } \label{bias_bis_percentage}
\end{figure}

We characterise the convergence of the iterative procedure by means of the relative residuals $\sum_{\kappa}|T_{i}^{\alpha}(k_{\kappa}) -1 |/N_{\rm F}$, where the sum is done over the total number of spherical shells in Fourier space $N_{\rm F}$ used to measure the different power spectra (i.e, from the fundamental mode to the Nyquist frequency). Fig.~\ref{bam_iterations} shows the behaviour of these residuals as a function of the number of iterations implemented for calibrations performed using two arbitrarily selected realisations of the Minerva simulations. Residuals of $\sim 1\%$ are achieved with approximately $\sim 30$ iterations, decreasing to a constant value of $\sim 0.9\%$ for $\sim 50$ iterations. We have adopted the $1\%$ threshold to stop the iterative process and generate \texttt{BAM} outputs required in the mock-construction.


\subsection{Production of mock density fields}\label{sec:prod}
The right-hand side panel of Fig.~\ref{bam_scheme} shows the steps followed to generate a mock halo density field based on the outputs from the calibration procedure of \texttt{BAM}, using a new set of realisations of the \texttt{IC}s. We have generated $N_{\rm sim}$ realisations of an \texttt{A-DMDF} $\delta_{\rm dm}^{j}$ ($j=1,\cdots, N_{\rm sim})$, using the $N_{\rm sim}$ down-sampled versions of the Minerva ICs. These density fields are convolved with the \texttt{BAM}-kernel, $\tilde{\delta}_{\rm dm}^{j(i\,\rm ref)}(\vr)\equiv \lp \mathcal{K}_{i}\otimes \delta_{\rm dm}^{j}\rp (\vr)$, after which the non-local properties (e.g cosmic-web types) of the resulting field $\tilde{\delta}_{\rm dm}^{i(j)}$ are determined. According to these properties, \texttt{BAM} statistically populates these fields with a number of haloes as
\begin{equation} 
\label{eq:mock_crea}
\{ N_{h}^{j(i\,\rm ref)}(\vr) \} \curvearrowleft 
\mathcal{B}\lp N_{\rm h}^{i\,{\rm ref}}\, \mid \, \{ \Theta_{\rm dm}^{i}\} _{\ell}= \{ \Theta_{\rm dm}^{j(i\,\rm ref)}(\vr)\}_{\ell}\rp,
\end{equation}
where the notation $j(i\,\rm ref)$ specifies that the $j$-th halo density field is constructed from the calibration obtained with the $i$-th realisation of the reference simulation. In \S~\ref{sec:results} we examine the statistical properties of an ensemble of realizations of halo number counts constructed following Eq.~(\ref{eq:mock_crea}), assessing their accuracy by comparing them against those of the reference simulation. 
\begin{figure*}
\hspace{-0.6cm}
\includegraphics[width=19cm]{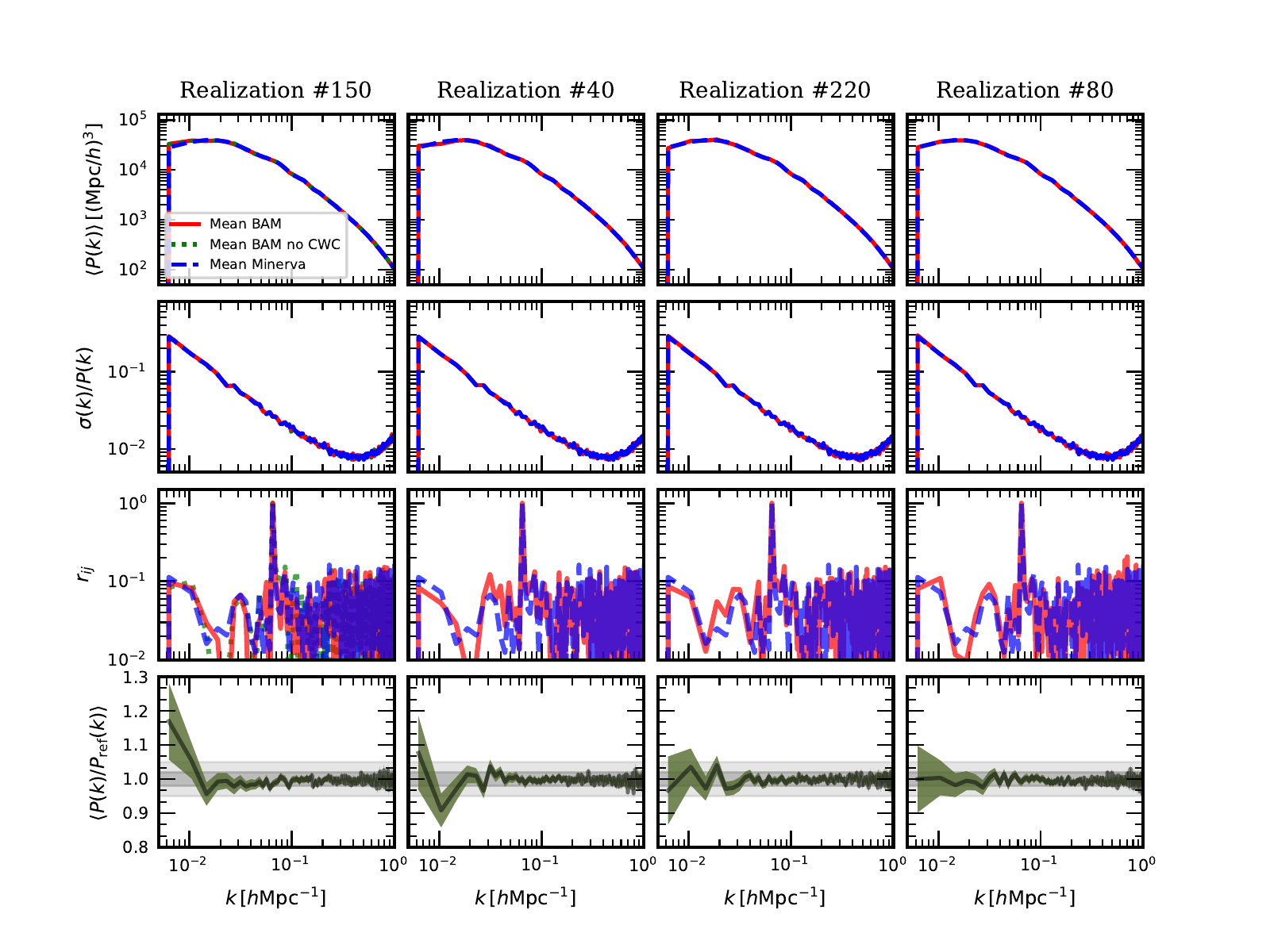}
\vspace{-1.cm}
\caption{Statistical properties in Fourier space of an ensemble of $300$ mock catalogues generated with \texttt{BAM} using \emph{one} reference realisation of the Minerva simulation with their respective \texttt{DMDF} evolved according to the Minerva \texttt{IC}s and \texttt{ALPT}. The different columns show examples of calibrations performed with different references, as highlighted in each plot-title. From top to bottom we present the mean power spectrum, its standard deviation $\sigma$ in units of the mean power spectrum, one element of the correlation matrix, and the average of the ratio between each realisation to the same realisation in the reference. The solid lines represent the different quantities computed from the \texttt{BAM} mock catalogues. Dotted lines represent the corresponding statistics from the full ensemble of the Minerva simulations. For the calibration performed with the realisation $\#150$ (first column) we show the results of generating mock catalogues without using cosmic-web classification (dotted line). The last row shows the ratio between the power spectrum of \texttt{BAM}  realisations to the corresponding (according to the \texttt{IC}) power spectrum of the reference ensemble, together with its variance (green-shaded areas), aimed at assessing the impact of populating different dark matter density fields using the bias and the \texttt{BAM} kernel calibrated from one reference. The shaded areas denote the $2\%$ and $5\%$ differences with respect to unity. This shows that the power spectra are within 2\% from $k\sim0.04 \,h$ Mpc$^{-1}$. However, the impact of cosmic variance can be seen in untreated kernels at lower $k$.} 
\label{bam_power}
\end{figure*}

Figure~\ref{bam_joints} displays contours from some of the joint distributions occurring within \texttt{BAM}. Panel (a) shows the halo-bias relation using a high resolution \texttt{DMDF} with its corresponding \texttt{DMH} catalogue. Panel (b) shows the same situation, this time using the \texttt{A-DMDF} generated by \texttt{ALPT}. The stochastic and non-linear behaviour of the halo-bias is evidenced in these two panels, the difference in their shapes being determined by the different \texttt{DM} distributions originated from each gravity solver. In panels (c) and (d) we show how the kernel modifies the 1D-distribution of the \texttt{DMDF}, making it broader towards bins of low \texttt{DM} density, effectively generating a \texttt{DMDF} with a lower clustering amplitude than that from the reference \texttt{DMDF}. Finally, the link between the initial \texttt{DMDF}s and their convolution with \texttt{BAM} kernel is presented in panels (e) and (f) for the reference and the \texttt{A-DMDF}, respectively. 
Interestingly, the kernel convolution straightens the bias relation, as can be seen comparing the  upper panels to the middle ones. Although the scatter in the bias relation is apparently enlarged, we will show below that the covariance matrices are well reproduced. Another key message from Fig.~\ref{bam_joints} is that the necessity of the iterative process within \texttt{BAM} is not simply due to the approximated nature of the gravity solver used to generate the \texttt{DMDF} to sample from, but also due to the incomplete picture of the halo bias as measured by Eq.~(\ref{eq:bias}). 

\begin{figure}
\includegraphics[width=9cm]{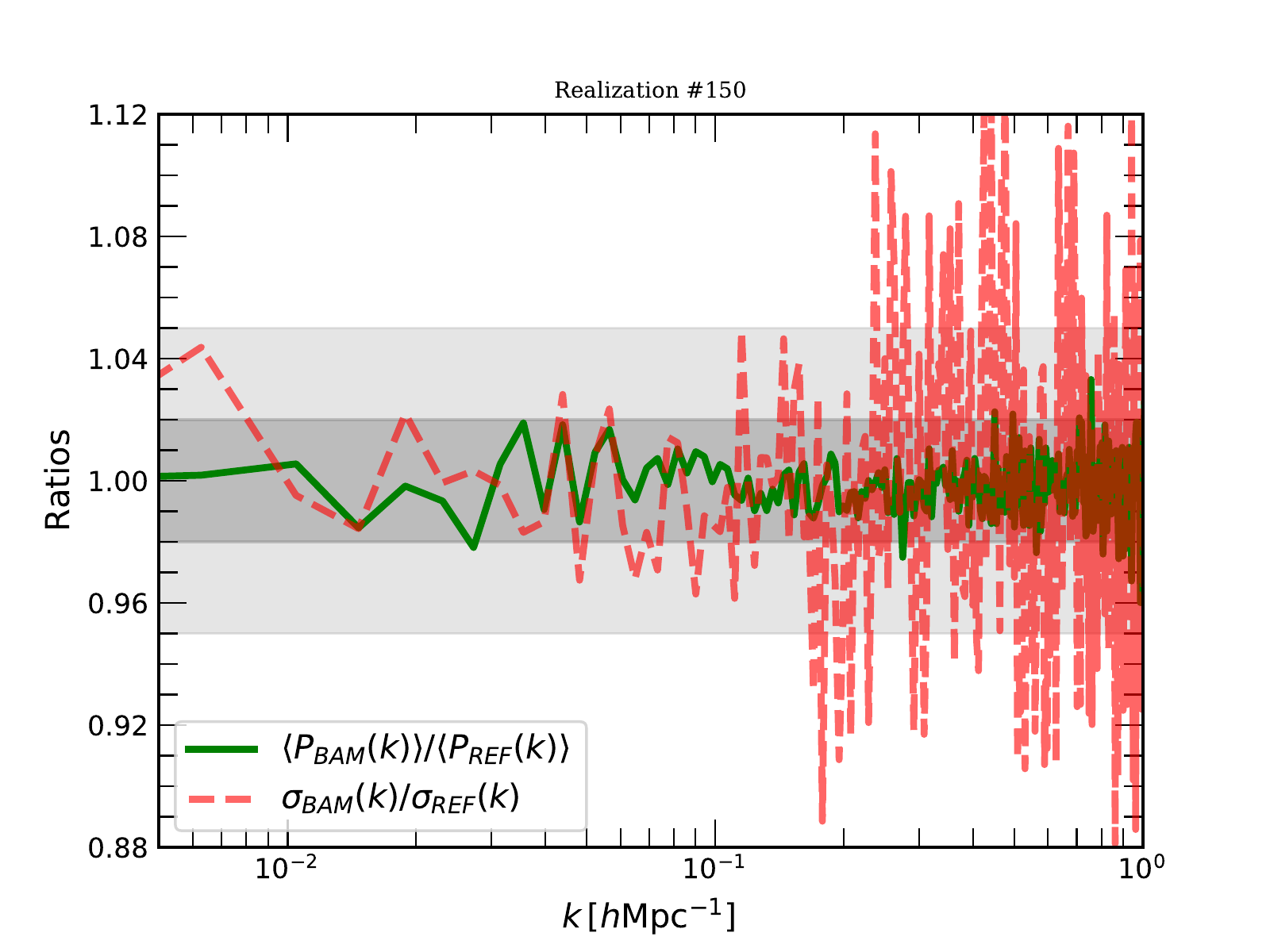}
\caption{Ratio of the mean \texttt{BAM} power spectrum and its variance from the set $j(150)$ to the same quantities obtained from the reference set of $N$-body simulations. The grey shaded areas denote the $2\%$ and $5\%$ deviations from unity.} 
\label{bam_power_onereal}
\end{figure}

In our approach we need to distinguish among different contributions to the stochasticity in the bias relation. One comes from our effective bias description, relating number counts of objects to the underlying dark matter field on a mesh. This small scale variance can be seen in the scatter of the bias relation in Fig.~\ref{bam_joints}. This contribution is fully accounted for in \texttt{BAM}, as it takes the whole bias relation extracted from reference simulations to assign number counts of objects per cell.
Another contribution comes from cosmic variance, as the reference simulation has a limited volume (or equivalently there is a limited number of reference simulations). In this regard, we have verified, that the bias relation in the setting of the reference catalogs is well described by a single simulation. This can be seen in Fig.~\ref{bias_bis_percentage}, where we show an statistical comparison between the bias $\mathcal{B}(N_{\rm h}, \delta_{\rm dm})$ (i.e, the joint probability of having $N_{\rm h}$ halos in a cell with a dark matter density $\delta_{\rm dm}$) measured from two realisations of the ensemble  $300$ Minerva halo catalogs $+300$ \texttt{A-DMDF}, against the mean of the same quantity measured from that reference ensemble. Within the region enclosing $\sim 98\%$ of the used cells, the measurements from the two examples of \texttt{BAM} mocks display differences with respect to the mean of the ensemble of the order of $5\%$, thus indicating that the effect of cosmic variance in the measurements of the halo bias are likely to be sub-dominant when obtained from a \texttt{BAM} calibration procedure.

Another impact of cosmic variance comes from the spatial distribution of halo number counts, causing noticeable (potentially larger than $10 \%$) deviations from the theoretical mean power spectrum on large scales.
This is taken into account within \texttt{BAM} in two steps. The first step, is based on using the same initial conditions (white noise) on large scales, as in the reference simulation.
The second step consists on a proper determination of the kernel, as we  will study in the next section. 

\section{\texttt{BAM} against N-body simulations}\label{sec:results}

In this section we assess the statistical properties of the ensemble of mocks generated with \texttt{BAM} and compare such properties against those obtained from the reference simulations in Fourier space. In order to demonstrate that \texttt{BAM} is not tailored for any particular realisation of the reference ensemble, we have arbitrarily selected four realisations of \texttt{IC}s and \texttt{HDF}s from the latter, viz, $i=40,80,150$ and $i=220$. For each one of these, the kernel $\mathcal{K}$ and the bias $\mathcal{B}$ are calibrated as described in \S~\ref{sec:calibration}, and an ensemble of $N_{\rm sim}=300$ \texttt{BAM} realisations of \texttt{HDF}s  are generated as explained in \S~\ref{sec:prod}.

\subsection{Power spectrum}\label{subsec:power}
We first investigate the power spectra from a set of four ensembles of \texttt{HDF} constructed using the calibration performed from the same number of reference simulations, as explained in \S~\ref{sec:prod}. Figure~\ref{bam_power} shows from the first to the third row the ensemble-mean power spectrum, the standard deviation (\texttt{SD}) of the mean and one element of the correlation matrix of the power spectrum, respectively. 
 As can be read from this plot, not only the mean of the \texttt{BAM} ensembles are in excellent agreement with the reference, but also the variance and the correlation matrix are in equally good correspondence. Indeed, the relative residuals (computed as shown in \S~\ref{sec:calibration}) are $\lesssim 1\%$ and $\lesssim10\%$ for the mean and the standard deviation, respectively. Figure~\ref{bam_power} then shows one side that the statistical error budget in the mock power spectra generated with \texttt{BAM} agrees with that of the reference, and that such performance is not due to the implementation of any particular reference simulation. 

The bottom panels of Fig.~\ref{bam_power} show, with green shaded areas, the standard deviation $(\sigma_{i})$ of the ratio between the $j$-th power spectrum in a \texttt{BAM} set $j(i)$ (i.e, calibrated with the reference $i$) against the spectrum of the reference associated to the same \texttt{IC}s. This ratio highlights the differences between power spectra of \texttt{HDF}s which, although originated from different gravity solvers ($N$-body and \texttt{ALPT}) and halo sampling (FoF and \texttt{BAM}), share the same initial conditions, and its deviations from unity accounts for the systematic effect induced by the outputs of the calibration. For the particular set of realisations shown in Fig.~\ref{bam_power}, this ratio shows values well compatible with unity, except for the largest scales where deviations of $1-2\sigma_{i}$ are observed, while displaying $\sim 5\%$ random fluctuations on small scales.  These deviations are nevertheless below the variance of the ratio of individual realisations to the mean.

Figure \ref{bam_power_onereal} presents a detailed view of the ratios between the mean power spectrum (and its variance) from from the set $j(150)$ against the same quantities obtained from the reference simulation. Deviations of the order of $\sim 1-2 \%$ in the mean power spectrum are observed up the the Nyquist frequency, while the variances vary from $\sim 2\%$ on large scales to to $10\%$ towards small scales.

\begin{figure}
  \includegraphics[width=9cm]{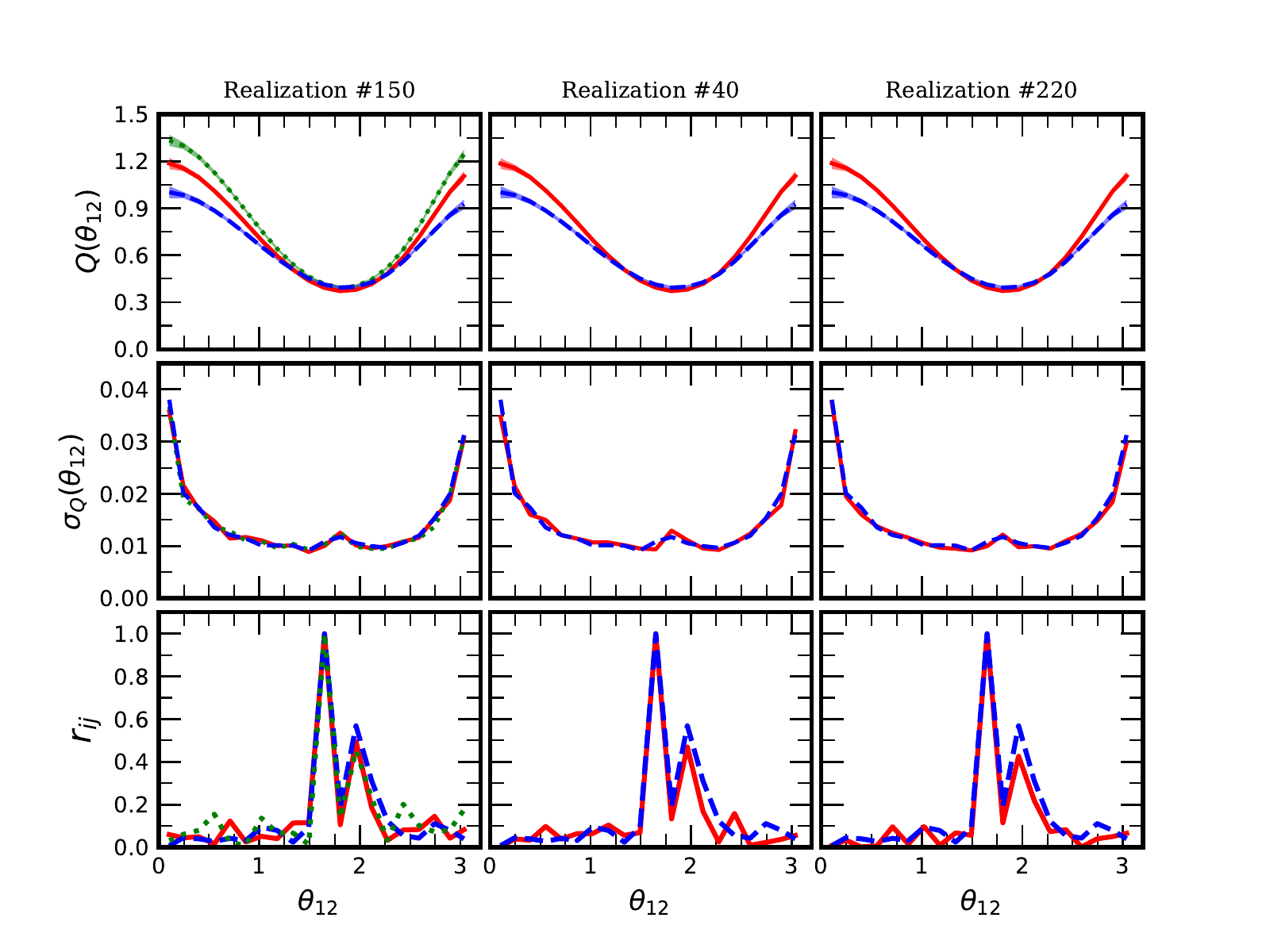}
\vspace{-0.5cm}
\caption{Statistical properties of the \texttt{BAM} mocks in Fourier space in terms of the reduced bispectrum for three different calibrations. In all panels, the blue-dotted (red-solid lines) lines represents the result from the reference (\texttt{BAM}) simulations. The top panels show the mean and variance (represented by shaded areas) of the signal for the \texttt{BAM} mocks and the reference. Second and third row show respectively the ensemble standard deviation and one example of the correlation matrix. The first column also shows the results of calibrating the \texttt{BAM} kernel and using a bias which is assumed to be dependent only on the local dark matter density.} 
\label{fig:bias_bis}
\end{figure}

\subsection{Bispectrum}\label{sec:bispectrum}
We now focus on the performance of our set of mocks in terms of the three-point statistics. In Fig.~\ref{fig:bias_bis} we show the reduced bispectrum $Q(\theta_{12})$ of the \texttt{BAM} mocks, as a function of the angle $\theta_{12}$ between the vectors $\vk_{1}$ and $\vk_{2}$ with $|\vk_{2}|=2|\vk_{1}|=0.2\,h$ Mpc$^{-1}$. The difference between the signal from the \texttt{BAM} ensemble (red shaded region) and that of the reference (blue-shaded regions) is evident, and amounts to $\sim 5$ times the variance. In principle, such discrepancy is likely to be due to unknown dependencies in the halo bias not accounted for in the calibration procedure. This conclusion is motivated by checking the results of running \texttt{BAM} assuming a halo-bias to be only a function of the local dark matter density. The first column of Fig.~\ref{fig:bias_bis} shows the statistics from the set of \texttt{BAM} realisations calibrated without any cosmic-web classification nor environmental dependencies. By construction, the calibration of the power spectrum is not sensitive to the lack of a given set of properties in the characterisation of the halo-bias, since the impact thereof is observed by the kernel. However, the effects of such neglected dependencies in the three-point statistics (which is not calibrated by \texttt{BAM}) are evident, as can be seen in the first column of Fig.~\ref{fig:bias_bis}, where we show the mean bispectra obtained without taking into account any cosmic-web classification (green shaded area). The impact of not considering the cosmic-web information increases the difference with respect to the reference from $\sim 4$ to $\sim 7$ times the variance. This $\sim 3\sigma$ of difference between the signal with and without cosmic-web information, can be regarded as a signature of non-local bias in the three point statistics, a topic we shall cover in a forthcoming work. It is also interesting that the sample variance in the two cases is in good agreement with that from the reference, as well as the element of the correlation matrix shown. One possible strategy to reproduce the correct three-point signal is to relax the definition of the cosmic-web types by using different values of the threshold \citep[e.g.][]{2009MNRAS.396.1815F}, or to increase the dimensionality of our bias measurement in order capture more information of the dark matter density field including e.g. the information of the velocity field. Although no improvement has been reached with the former, we are currently working in including the latter. 

In \citet{2019arXiv191013164P} we have shown that this discrepancy is also sensitive to the mean number density of the sample (and hence the shot-noise level), significantly reducing when applying the method to higher resolution reference simulations. 
\subsection{Covariance matrices}\label{sec:like}
In this section we compare the behaviour of the covariance matrix of the power spectrum from the different sets of halo mock catalogues with respect to that from the reference one. We have chosen the \texttt{BAM} sets $j(40)$, $j(150)$ and $j(220)$, and perform fits to the parameters of a given power spectrum model in the likelihood analysis framework. This test goes in the spirit of recent efforts to compare the performance of different approximate methods to generate mock catalogues \citep{2019MNRAS.482.1786L,2018arXiv180609497B,2018MNRAS.tmp.2818C}.

\begin{figure}
\hspace{-0.5cm}
 \includegraphics[width=0.5\textwidth]{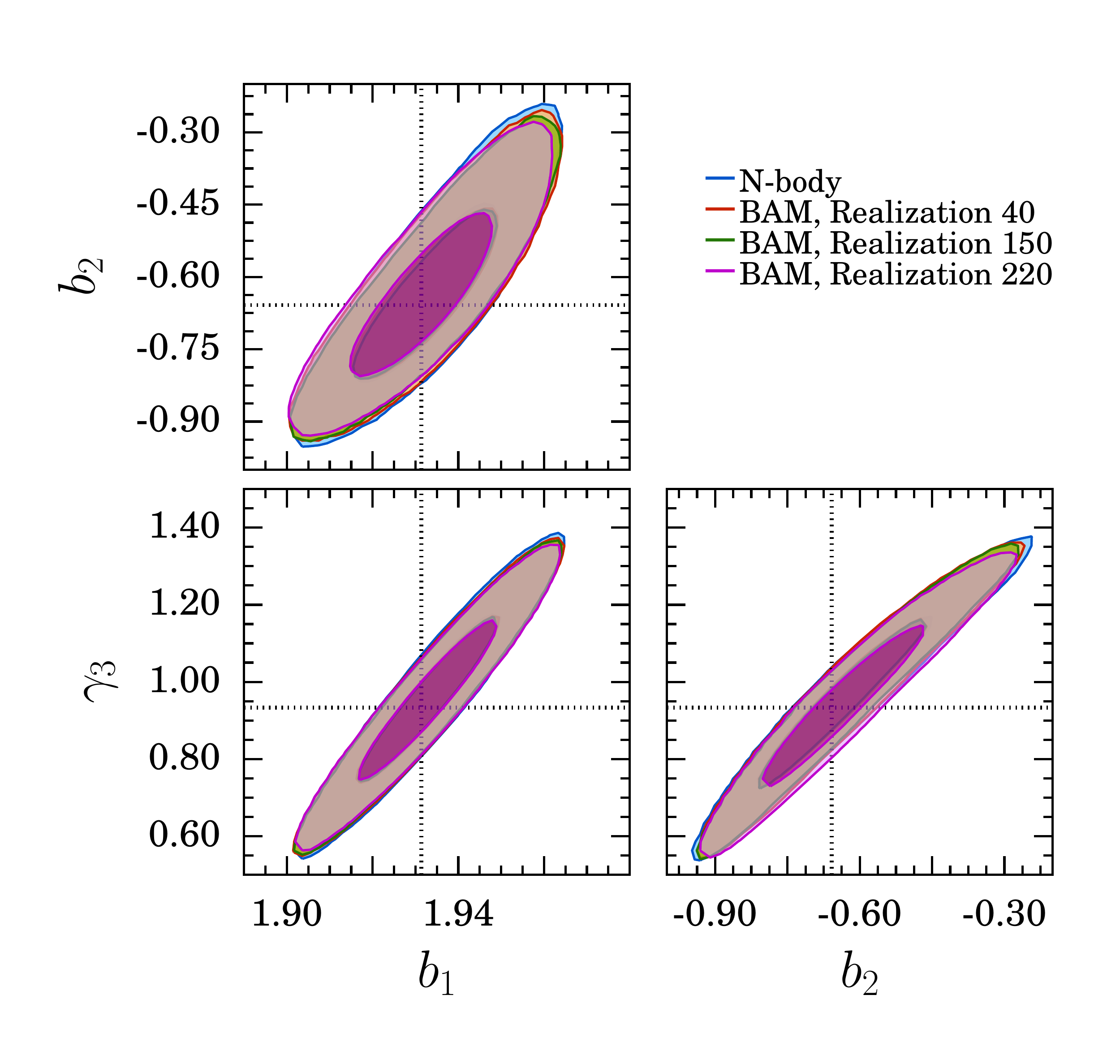}
\hspace{-0.8cm}
\vspace{-0.5cm}
 \caption{Comparison of the marginalised two-dimensional constraints on the nuisance parameters obtained from the analysis using the covariance matrices from the different \texttt{BAM} catalogues against that from the reference simulations. The contours correspond to the $68\%$ and $95\%$ confidence levels.
 The dotted lines indicate the expected mean parameter values.}
 \label{fig:2dconstraints_nbody}
\end{figure}

The covariance matrices are computed from the different sets of $N_s$ mock catalogues as $C_{ij}=\langle P^{k}_{i}-\langle P_{i}\rangle_{k} \rangle \langle P^{k}_{j}-\langle P_{j}\rangle_{k} \rangle_{k} $
 where $\langle P_{i}\rangle_{k}$ 
is the mean value of the measurements at the $i$-th bin and $P_i ^k$ is the corresponding measurement from the $k$-th mock. We take a similar approach to \citet{2018arXiv180609497B, 2019MNRAS.482.1786L}, with the only difference that we focus on the real-space power spectrum covariance, not including redshift-space distortions.
We adopt the same model for the real-space power spectrum, which is based on re-normalised perturbation theory \cite[see e.g.][]{2017MNRAS.464.1640S}, containing three free parameters, viz, the linear bias $b_1$, the quadratic bias $b_2$ and the non-local bias parameter $\gamma_3^-$. All cosmological parameters are kept to their true values, and we use Fourier modes in the range of $0.008 < k/(\,h\,{\rm Mpc}^{-1}) < 0.25$. 
We construct synthetic power-spectrum data using the parameters obtained from the fit of the mean power spectrum measured from the $N$-body reference halo catalogues with the measured reference covariance matrix and our baseline model. This ensures that our data is perfectly described by our baseline model and differences in the obtained constraints depend only on the covariance matrices. 
We perform MCMC fits of the theory power spectrum using the four different covariance matrices and assuming a Gaussian likelihood.

\begin{figure}
\hspace{0.5cm}
 \includegraphics[width=0.4\textwidth]{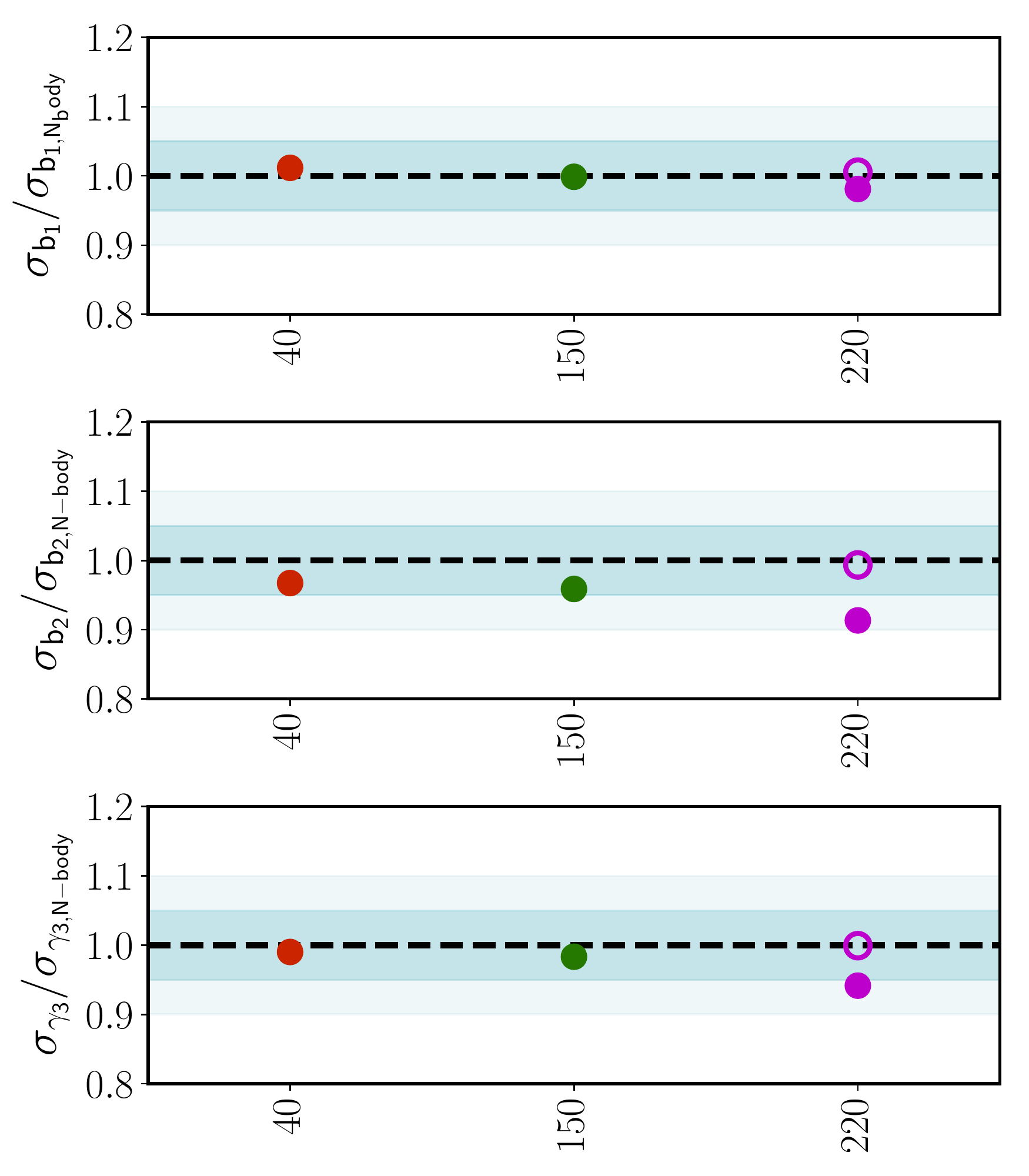}
\hspace{-0.5cm}
 \caption{Comparison of the marginalised one-dimensional parameter errors obtained from the analysis using the covariance matrices from the different \texttt{BAM} catalogues against that from the reference simulations. The light grey bands indicate ranges of $\pm 5 \%$ and $\pm 10 \%$. The empty dot show the errors obtained from the set $j(220)$ constructed with a kernel obtained from the mean of $100$ independent calibrations, as discussed in \S~\ref{sec:like}. }
 \label{fig:paramerrors}
\end{figure}

Figure \ref{fig:2dconstraints_nbody} shows the two-dimensional marginalised constraints obtained from the analysis with the reference and the three different \texttt{BAM} covariance matrices. We find that the mean parameters resulting from the analysis using the three different \texttt{BAM} covariance matrices are in excellent agreement with the expected parameter values. The resulting 2$\sigma$-parameter contours are also well reproduced by the \texttt{BAM} covariance matrices, only for some cases very slightly under- or over- estimating the statistical uncertainties. Therefore, we also consider the one-dimensional marginalised parameter errors drawn from the analysis with the different \texttt{BAM} covariance matrices. Figure~\ref{fig:paramerrors} shows the ratios of the parameter errors from the results obtained with the \texttt{BAM} mocks with respect to those obtained with the reference. The three \texttt{BAM} covariance matrices reproduce the $N$-body errors within $10\%$, in most cases even within $5\%$, corresponding to the expected statistical limit of our analysis \citep[see][]{2018arXiv180609497B}. 

We have also constrained the set of parameters for the power spectrum model using the covariance matrix from an ensemble of mock catalogs built with a mean kernel obtained from the calibration of $100$ independent kernels. The blank symbols in Fig.~\ref{fig:paramerrors} show the results obtained from the set built with the halo-bias from reference $\#220$. This approach generates excellent results, with errors which differ $\sim 1\%$ with those obtained from the reference. We note however that the same procedure is not guaranteed to yield the same results when using the bias calibrated from another realisation. The reason is that any incompatibility between the kernel and the bias is propagated into the error analysis to the off-diagonal elements of the covariance matrix. In other words, the construction of mock catalogs within our method has to be performed with a halo-bias obtained from the same reference used to calibrate the kernel, if precise and accurate covariance matrices are expected to be achieved. The example shown in Fig.~\ref{fig:paramerrors} for the set $j(220)$ generates good results due to the fortuitous fact that the kernel from the realisation $\#220$ is in good agreement with the mean kernel used to construct that set of mocks. 

We have verified that on large scales, the mean kernel is smooth and displays a constant amplitude. While the former property (expected from an average of $100$ independent kernels) contributes at reducing the large-scale discrepancies observed in the bottom panels of Fig.~\ref{bam_power}, the latter opens the possibility to construct mock catalogues with volumes larger than that of the reference simulation, by extrapolating the constant kernel to the desired cosmological volume. This however, involves the introduction of large-scale modes, not present in the reference simulation, which might affect the accuracy of the \texttt{BAM} calibration. We shall assess this effect in forthcoming publications.

Although these conclusions are obtained with $100$ independent calibrations, we stress that the same goal can be achieved by mean of the concept of paired-fixed simulations \citep[][]{2016MNRAS.462L...1A,2018ApJ...867..137V,2018arXiv181102111C}, consisting in a set of \emph{two} reference simulations generated with the same initial conditions, with fixed amplitude of initial power spectra and inverted phases. As an example, we have verified that using the set described by \citep[][]{2018arXiv181102111C} we obtain a smooth kernel with constant amplitude on large scales. With this approach we also obtain halo-bias corresponding to such kernel, thus providing the perfect scenario for the good performance of the covariance matrices, as we will show in a forthcoming publication.


\section{Discussion and conclusions}
In this paper we have assessed the accuracy of the recently published \texttt{B}ias \texttt{A}ssignment \texttt{M}ethod (\texttt{BAM}), described in \citet[][]{2019MNRAS.483L..58B} (Paper I). To that aim we have used a reference $N$-body simulation (the Minerva simulation, \citet[][]{2016MNRAS.457.1577G}) and \emph{one} arbitrary member (out of its $300$ realisations) in order to calibrate the main outputs of our method. Such outputs (viz, a halo-bias and a kernel  in Equations (\ref{eq:bias}) and (\ref{eq:kernel}) respectively) are used to sample $300$ dark matter density fields obtained from the evolution of the initial conditions of the reference simulation using Augmented Lagrangian Perturbation Theory (\texttt{ALPT}). This procedure has been repeated using other realisations as references, in order to demonstrate that our method is not tailored for any particular realisation. The halo mock catalogues produced with \texttt{BAM} display power spectra which reproduce the reference ones with $2\%$ precision at $k\sim 0.5 h\, {\rm Mpc}^{-1}$ up to $5\%$ precision at the Nyquist frequency ($k\sim 1\, h {\rm Mpc}^{-1}$), while the mean of the ensemble agrees to $\sim 2\%$ with that from the reference ensemble up to the Nyquist frequency (see Fig.~\ref{bam_power_onereal}). We have verified that covariance matrices of the two and three point statistics measured from the mock catalogues generated by \texttt{BAM} are in very good agreement with those obtained from the reference $N$-body simulation. In terms of the power spectrum, such agreement with the reference is translated into a $\sim 5\%$ difference between the set of uncertainties in parameters of models of power spectrum. Furthermore, we have shown that the implementation of an smooth (on large scales) version of the kernel in conjunction with a compatible bias (i.e, a bias that is calibrated with the kernel from the same reference simulation) can further reduce the differences in the errors to $\sim 1\%$ with respect to those obtained with reference covariance matrix. 


The \texttt{BAM} approach is based on the representation of all the non-linear and non-local bias relations in an explicit and parameter free way, extracted from the information of one $N-$body simulation. In particular, we extract the local and non local properties of the approximated dark matter density field constructed with \texttt{ALPT}.
Among such non-local quantities we consider the eigenvalues of the tidal field tensor and the total mass of the collapsed regions (see Paper I). We are fully aware that a series of additional non-local relations are not being accounted for in our description, such as the effective kernel describing a halo finder with respect to our dark matter field representation (in particular the mass assignment kernel of a cloud-in-cell operation) and the missing power of Lagrangian perturbation theory based methods \cite[see e.g.][]{2017JCAP...07..050M}. But even using more accurate gravity solvers, such as \texttt{FastPm} or \texttt{COLA}, as we are using a very low number of particles to describe the dark matter field (for the sake of reducing the computational requirements), we are prone to have some systematic deviations with respect to an accurate representation of the dark matter field, especially towards small scales. Additionally, there are some potential contributions from non-local bias terms which we are not modelling, such as third order contributions \citep[see e.g.][]{2009JCAP...08..020M}. Even our second order non-local bias description could still leave some effective aliasing towards small scales, due to, for instance, the particular computation of the eigenvalues (grid resolution, threshold, etc). Since the resulting effective kernel becomes quite complicated, and is unknown in our case, being the sum of all above mentioned effects, we are estimating it iteratively to minimize the deviation from the reference power spectrum. In this way we leave the remainder of the effects we are not explicitly modelling as part of the process involved in the calibration our so-called \texttt{BAM} kernel. 

Our method proposes to use \emph{one simulation to have them all}. One such simulation (i.e, the reference) is intended to represent a sufficiently large cosmological volume with dark matter tracers resolved at a high-mass resolution. This allows for a precise calibration (with a high signal-to-noise  ratio in the number counts-per-cells) and consequent construction of mock catalogs with applications to galaxy surveys characterised by different estimates of minimum halo masses. For the present work, the Minerva simulation offers the scenario in which the (Poisson) signal-to-noise ratio in the halo number counts-per-cells ($1$ to $3$ for a spatial resolution of $3$Mpc $h^{-1}$) allows for a significant determination of the halo-bias which in turn permits the convergence of the iterative procedure described in \S~\ref{sec:meth}. Nevertheless, the impact of the shot-noise is likely to be propagated into the higher order statistics, partially accounting for the behavior shown in Fig.~\ref{fig:bias_bis}. In \citet[][]{2019arXiv191013164P} we have shown that higher signal-to-noise ratios in the counts-per-cells lead to much more precise higher order (e.g. three-point) statistics in the \texttt{BAM} mocks.

We note that the version of \texttt{BAM} presented in this work provides catalogs in the form of number counts; individual halo properties (e.g. comoving coordinates, masses, velocities) are to be assigned following procedures such as that suggested by \citet[][]{2015MNRAS.451.4266Z}. Once these properties are assigned, similar tests as the one presented in this paper can be performed, focused on the impact of cosmological parameters, as has been done in e.g. \citet[][]{2019MNRAS.485.2806B, 2019MNRAS.482.1786L}.

The iterative procedure designed to determine the kernel can be interpreted as a particular case of a machine learning method in which  the \emph{cost function} is represented by the power spectrum. 
Usual deep learning approaches require in general as a training set a large number of catalogues from detailed reference simulations  \citep[see e.g.][]{2017arXiv170602390M,2018arXiv180804728M,2018arXiv181106533H}. Such an approach would then capture the nonlinear and non-local effective relation between the halo and the dark matter field. This represents a problem, given that $N$-body simulations addressing the current galaxy survey requirements are highly demanding (e.g. the Euclid Flagship simulation\footnote{http://sci.esa.int/euclid/59348-euclid-flagship-mock-galaxy-catalogue/}). One can certainly go for smaller volumes with techniques such as that suggested by \citet[][]{2016MNRAS.462L...1A}, but even those are very demanding, and only a few can be done \cite[see e.g.][]{2018arXiv181102111C}.
In this work, we use as much information as we have on the physical and statistical relation between the halo and the dark matter distribution to alleviate the need for large training sets.

A second problem of the application of the concepts of machine learning to large scale structure comes from defining the appropriate \emph{cost function}. In this context it is in general dangerous to define as a cost function only the power spectrum, as the bias relation is degenerated at the three-point statistics level \cite[see][]{2015MNRAS.450.1836K, 2017MNRAS.472.4144V}. See, however some interesting recent work using the number counts as the cost function to learn from a single simulation  \citep[e.g.][]{2019arXiv190205965Z}, as we essentially proposed in Paper I. In \texttt{BAM} we can afford restricting our cost function to second order statistics because we do not leave any freedom to the bias relation, which is fully extracted from the reference simulation, as explained above. 

Our approach offers the clear advantage of demanding one (at most two) reference simulations as a training set, as well as allowing for the identification of quantities (e.g. the halo-bias and the kernel) encoding physical information (such as assembly bias, e.g.  \citet[][]{2007IJMPD..16..763Z,2007MNRAS.377L...5G,2008MNRAS.387..921A,2017JCAP...03..059L,2018JCAP...10..012C,2019MNRAS.484.1133C}) related to the formation and evolution of dark matter tracers. Instead of being competitors, the \texttt{BAM} approach and a machine learning technique can be complementary, as an example, at assigning intrinsic properties to the tracers in our mock catalogues, a task we are currently working on.

In summary, our approach extracts the nonlinear and non-local information from a reference simulation, leaving the remaining unknown contributions to a single linear kernel. 
The method succeeds because a single reference (large enough) simulation encodes a vast information about the bias relation and its dimensionality is effectively given by the number of cells for different density bins, tidal field tensor eigenvalues, and total mass of connected knots. We note, that we can extend our approach to account for third order bias contributions within the same framework, allowing us to improve the precision of the \texttt{BAM} mocks in terms of their three-points statistics. Still a number of issues have to be answered, such as including redshift space distortions, going to the mass resolution to resolve the haloes hosting emission line galaxies as required for surveys, such as Euclid or DESI, and populating those halos according to each survey selection function. Nevertheless, the present work shows that our method has the potential to generate precise and accurate mock catalogues based on high resolution $N-$body simulations.

\section*{Acknowledgements}
We acknowledge the anonymous referee for his/her valuable comments, which lead to considerably improve the presentation of our work.
ABA acknowledges financial support from the Spanish Ministry of Economy and Competitiveness (MINECO) under the Severo Ochoa program SEV-2015-0548. FSK thanks support from grants SEV-2015-0548, RYC2015-18693 and AYA2017-89891-P. MPI acknowledges support from MINECO under the grant AYA2012-39702-C02-01. MPI wishes to acknowledge the contribution of Teide High-Performance Computing facilities for the results of this research. TeideHPC facilities are provided by the Instituto Tecnol\'{o}gico y de Energ\'{i}as Renovables (ITER, SA). REA and MPI acknowledge the Spanish MINECO and the European Research Council through grant numbers  \textit{AYA2015-66211-C2-2} and \textit{ERC-StG/716151}, respectively.

\bibliographystyle{mnras}
\bibliography{refs}  
\end{document}

%% file: newcomands.tex
\newcommand{\be}{\begin{equation}}

\newcommand{\lp}{\left(}
\newcommand{\rp}{\right)}

\newcommand{\vr}{\textbf{r}}

\newcommand{\ee}{\end{equation}}

\newcommand{\vk}{\textbf{k}}

\newcommand{\vq}{\textbf{q}}